# Elastic Energy Storage Mechanism in Hovering Animal Flight: A Discriminative Method Based on Wing Kinematics

Shijie Sheng[1], Jianghao Wu[1], Renxuan Bo[1], Long Chen[1], Yanlai Zhang[1,*]

[1] School of Transportation Science and Engineering, Beihang University, Beijing, China
\* Author for correspondence: Yanlai Zhang, email: zhangyanlai@ buaa.edu.cn

## Abstract

Existing research has yet to reach a consensus on whether and how small flying animals utilize elastic energy storage mechanisms to reduce flight energy expenditure, and there is a lack of systematic and universal methods for assessment. To address these gaps, this study proposes a method to evaluate elastic energy storage capacity based on wing kinematic parameters (flapping amplitude and flapping frequency), grounded in the hypothesis that animals tend to minimize flight energy expenditure. By establishing a simplified power model, the study calculates the optimal kinematic parameters corresponding to the minimum mechanical power requirements under two extreme conditions: "no elastic energy storage" and "complete elastic energy storage". These optimal parameters are then compared with measured data from various small flying animals. The results show that the measured parameters of hummingbirds, ladybugs, and rhinoceros beetles are close to the no-storage optimum, indicating relatively weak elastic energy storage capacity; whereas hoverflies, bumblebees, and honeybees align closely with the complete-storage optimum, suggesting strong elastic energy storage ability. Furthermore, the wing kinematic adjustment strategies these animals employ in response to changes in load or air density are consistent with the predicted elastic storage capacities. This study provides a systematic new approach for assessing biological elastic energy storage capacity and offers a theoretical basis for the low-power design of flapping wing micro air vehicles.

## INTRODUCTION

Low energy expenditure is a key factor enabling both biological and man-made aircraft to achieve prolonged flight. Unlike man-made fixed-wing or rotary-wing aircraft, flying animals typically rely on reciprocating flapping wings to generate lift and thrust. This reciprocating flapping motion requires additional energy to overcome the inertial power requirement of the wings. Previous studies have shown that certain physiological structures in flying animals—such as tendons and resilin—possess elastic energy storage capabilities. These structures may reduce the mechanical power requirement for wing flapping by alternately storing and releasing part of wing's kinetic energy, thereby lowering overall flight energy expenditure (*1–3*). However, when studying the forward flight of animals such as birds, laboratory observations struggle to replicate the full range of flight conditions, while field observations are easily affected by gusts or external energy inputs (for example, seagulls exploit rising air currents to achieve long-duration, low-energy gliding). These factors present challenges in accurately assessing the energy expenditure of biological flight and uncovering their elastic energy storage mechanisms. In contrast, the hovering flight mode is not only easier to observe in the laboratory but is also largely unaffected by external energy inputs, making it an ideal model for studying biological flight energy expenditure and elastic energy storage mechanisms.

In nature, hovering is the primary mode of flight for small flying animals, such as hummingbirds and insects. This form of flight, which requires minimal external energy input and is easy to observe, greatly facilitates the study of whether these animals use elastic energy storage mechanisms to reduce the energy expenditure of flight. Experimental observations (*4, 5*) indicate that these species engage in rapid forward flight or maneuvering only when responding to specific needs, such as escaping, hunting, or avoiding obstacles. For the majority of their flight time, they maintain hovering or slow forward flight—states that involve almost no external energy input. Therefore, the energy expenditure during hovering or slow forward flight constitutes the main portion of their total flight energy expenditure. Additionally, studies have shown that the power requirement of these small flying creatures during hovering and slow forward flight does not differ significantly (*6–9*), making hovering flight a commonly chosen state for researching their flight power requirements and the potential existence of elastic energy storage mechanisms (*10–12*).

The mechanical power requirement for hovering flight in small flying animals consists of two components: the aerodynamic power required to overcome aerodynamic drag during the flapping cycle, and the inertial power required to accelerate and decelerate the wings (*13*), as illustrated in Fig. 1A. Notably, the mechanical power requirement differs between the wing acceleration and deceleration phases. During acceleration, the mechanical work performed by the flight muscles partially increases the wing's kinetic energy (overcoming inertia) and partially overcomes aerodynamic drag. During deceleration, the wing continues flapping due to its inertia, with the wing's kinetic energy partially or fully overcoming aerodynamic drag. Consequently, power requirements during deceleration are typically lower than during acceleration.

During the reciprocating acceleration and deceleration of the wing, recovery of excess kinetic energy can significantly reduce the total mechanical power requirements. When inertial power dominates, the kinetic energy gained by the wing during acceleration not only suffices to overcome aerodynamic drag during deceleration but also generates excess energy (Fig. 1B). If this excess kinetic energy is stored in an elastic element and subsequently released to drive wing acceleration, it can offset part of the mechanical power requirement in the next acceleration phase (*10*, *14*). Specifically, when excess kinetic energy is fully recovered, the mean inertial power over the entire flapping cycle becomes zero. In this scenario, small flying animals only need to supply aerodynamic power to sustain hovering flight. Conversely, if excess kinetic energy is dissipated as heat (or other forms) instead of being stored, these small flying animals must meet the combined requirements of aerodynamic and inertial power.

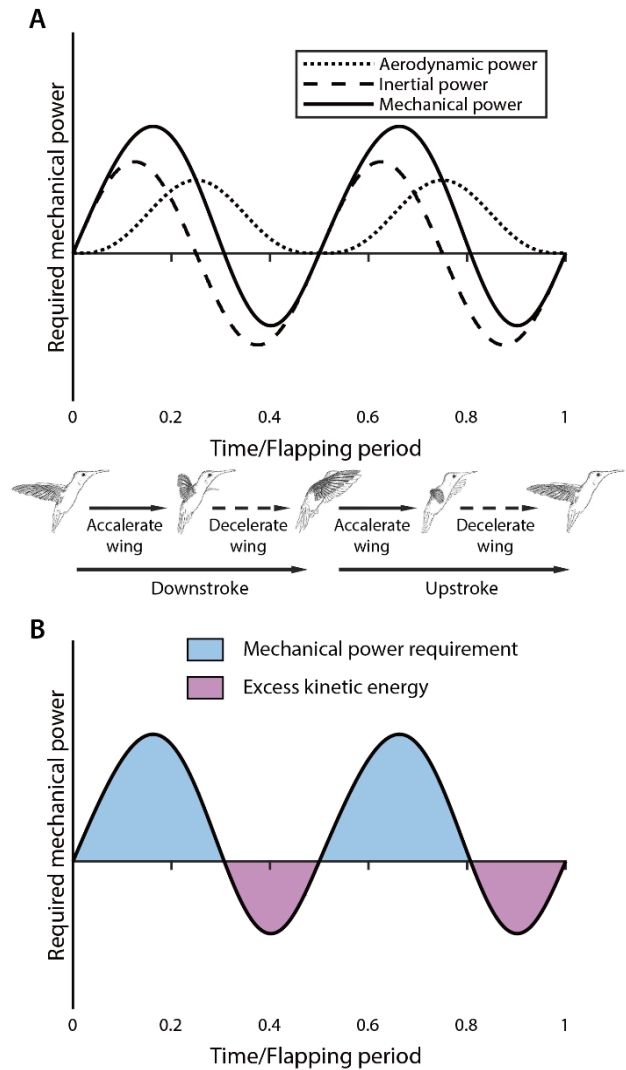

**Fig. 1. Mechanical power requirement for hovering flight in small flying animals.**
**(A)** Schematic of instantaneous aerodynamic power, inertial power, and mechanical power during one flapping cycle. Instantaneous mechanical power is the sum of instantaneous aerodynamic and inertial power. Throughout the cycle, aerodynamic power remains predominantly positive, while inertial power alternates between positive and negative twice, corresponding due to the wing's two acceleration and deceleration phases.
**(B)** When small flying animals lack the capability of elastic energy storage, the mechanical power requirement for hovering corresponds to the area shaded in blue in the figure, and the excess kinetic energy of the wings (represented by the purple shaded area) is dissipated. However, when these animals have complete elastic energy storage capability, the excess kinetic energy of the wings is stored in elastic structures and released during subsequent wing acceleration phases, allowing this energy to be fully utilized. In this case, the mechanical power requirement for hovering is the blue area minus the purple area. Compared to the scenario without elastic energy storage, the mechanical power requirement for hovering is significantly reduced with complete elastic energy storage.

Whether small flying animals utilize elastic energy storage mechanisms to reduce energy expenditure remains a subject of debate. Early studies (*1*, *13*, *15–18*) experimentally evaluated the mass-specific power (mechanical power/muscle mass) or efficiency (mechanical power/metabolic power) of flight muscles under two extreme conditions, complete elastic energy storage and no elastic energy storage, and compared these results with their theoretical limits. If the results assuming no elastic energy storage exceeded the theoretical limit, the animal must possess certain elastic energy storage capabilities. However, the results obtained using this method have been controversial. On one hand, the quasi-steady model (*12*), which is widely used in these studies to estimate the aerodynamic drag coefficient using $7/\sqrt{Re}$, may lead to a severe underestimation of aerodynamic power. This, in turn, results in an underestimation of the mass-specific power or efficiency of the flight muscles. For example, Song et al. (*19*) conducted a CFD study demonstrating that, under the assumption of complete elastic energy storage, the aerodynamic mass-specific power (aerodynamic power/body mass) of hummingbirds hovering can reach 55 W kg$^{-1}$, far exceeding Wells' (*17*) result (<30 W kg$^{-1}$). Similarly, Liu and Sun's (*20*) CFD study indicated that the aerodynamic mass-specific power of hoverflies during hovering ranges from 34.4 to 50.2 W kg$^{-1}$, which is also significantly higher than Ellington's (*12*) result (23.2–23.6 W kg$^{-1}$). On the other hand, the theoretical limit values of muscle mass-specific power—an important criterion for assessing elastic energy storage capacity—vary significantly across studies. For example, Weis-Fogh et al. (*21*) proposed a limit of 230–250 W kg$^{-1}$, Ellington (*16*) estimated 250 W kg$^{-1}$, Pennycuick and Rezende (*22*) suggested it could reach 430 W kg$^{-1}$, while Josephson (*23*) proposed a much lower value of only 100 W kg$^{-1}$. These factors have even led to completely opposite conclusions about the elastic energy storage capacity of the same species in different studies. For instance, Weis-Fogh (*1, 15*) believed hummingbirds lack elastic energy storage capability, whereas Wells (*17*) held opposite view. Similarly, Dickinson and Lighton (*13*) indicated that fruit flies have at least 11% energy storage capacity, while Lehmann(*24*) found that the aerodynamic power during fruit fly hovering exceeds inertial power, thus concluding that elastic energy storage is unnecessary for them.

Beyond investigations based on the power output capability of muscles, researchers have sought direct structural evidence of elastic energy storage. Reid et al. (*25*) found that the strain energy generated by wing deformation in the hawkmoth (*Manduca sexta*) can be recovered and reused to reduce the net energy expenditure of flight, accounting for approximately 30% of the total energy demand during its flapping cycle. Research by Gau et al. (*26*) demonstrated that the reduction in power requirements attributable to thoracic exoskeleton deformation in the hawkmoth ranges from 25% to 60% of the inertial power. Agrawal et al. (*27*) modeled the musculoskeletal system of hummingbirds and confirmed that, although their tendons have elastic energy storage capabilities—stronger than those of pigeons—they are still weaker than those of locusts and bumblebees.

Although existing studies have examined, from various perspectives, whether small flying animals use elastic energy storage mechanisms to reduce the energy expenditure of hovering flight, and research on physiological structures has provided some evidence, conclusions regarding certain species remain inconsistent. Moreover, a systematic and universal method to determine whether the flight of a particular species utilizes elastic energy storage mechanisms is still lacking. Therefore, this paper proposes a novel analytical approach: assessing elastic energy storage capability by analyzing wing kinematics during hovering. Because small flying animals expend considerable energy during hovering flight, they are likely to optimize wing kinematic parameters to reduce energy expenditure and enhance hovering efficiency. Therefore, we propose the following hypothesis: the wing kinematic parameters adopted by species with strong elastic storage capability should approach the optimal parameters under complete elastic storage conditions—those that minimize aerodynamic power. Conversely, parameters adopted by species with weak or absent elastic storage capability should approach the optimal parameters under no elastic storage conditions—those that minimize total mechanical power (i.e., the combined aerodynamic and inertial power). Aerodynamic power and inertial power are influenced by wing kinematic parameters—specifically flapping amplitude ($\Phi$) and flapping frequency ($n$)—through distinct scaling relationships (aerodynamic power scales with $\Phi^3 n^3$, whereas inertial power scales with $\Phi^2 n^3$ (*13*)). This results in distinguishable optimal wing kinematic patterns between the two extreme elastic storage scenarios. Consequently, elastic energy storage capability can be effectively assessed by quantifying deviations of measured wing kinematics from these two theoretical optima.

Based on the approach described above, we can assess the elastic energy storage capability of small flying animals as follows. Under the condition of a fixed measured flapping amplitude ($\Phi$) of a small flying animal, we varied its flapping frequency and calculated the hovering power requirements under two extreme elastic energy storage states. This approach allows us to identify two optimal flapping frequencies: $n_0$ (which minimizes hovering power with no elastic energy storage) and $n_1$ (which minimizes hovering power with complete elastic energy storage), where the subscripts "0" and "1" denote "no" and "complete" elastic energy storage, respectively. If the measured frequency ($n$) approaches $n_0$, it suggests weak elastic energy storage capability; if $n$ approaches $n_1$, it indicates strong capability; and if $n$ lies between $n_0$ and $n_1$, it implies partial utilization of excess kinetic energy during wing deceleration, reflecting moderate storage capability. Similarly, under the condition of a fixed measured flapping frequency ($n$), we calculated the optimal flapping amplitudes ($\Phi_0$, $\Phi_1$) that minimize hovering power requirements under the two extreme elastic energy storage states. By comparing the relative differences between these two amplitudes and the measured value ($\Phi$), the elastic energy storage capability could be determined. Throughout these computations, the mean angle of attack was adjusted to maintain lift-weight equilibrium as $n$ or $\Phi$ varies. The feasibility of this method has been validated by previous study (*28*). Based on the above methods, we analyzed hovering flight data from diverse small flying animals to evaluate their elastic storage capabilities.

# RESULTS

Based on previous studies (*10*, *12*, *17*, *20*, *29–38*), morphological data on diverse small flying animals were collected, along with their wing kinematic parameters during hovering or near-hovering conditions (i.e., flapping amplitude and flapping frequency), as detailed in Table S1. Subsequently, a simplified model was developed to rapidly assess the mechanical power requirements for flapping-wing hovering. By combining this model with the aforementioned parameters, the theoretically optimal wing kinematic parameters for these animals were calculated under two extreme energy storage conditions—namely, no elastic energy storage and complete elastic energy storage. These optimal parameters, denoted as $n_0$, $\Phi_0$ (corresponding to the minimum combined aerodynamic and inertial power) and $n_1$, $\Phi_1$ (corresponding to minimum aerodynamic power), are detailed in Table S2.

By comparing the relative differences between the measured parameters ($n$, $\Phi$) and the theoretical optimal parameters ($n_0$, $\Phi_0$, $n_1$, $\Phi_1$) (Fig. 2), the elastic energy storage capability of these animals can be assessed. The analysis indicates that, within the same species, the assessment results based on flapping frequency (Fig. 2A) and flapping amplitude (Fig. 2B) are highly consistent. Based on these findings, these small flying animals can be broadly categorized into two groups according to their elastic energy storage capability: the weak elastic energy storage group and the strong elastic energy storage group. The specific classification is as follows.

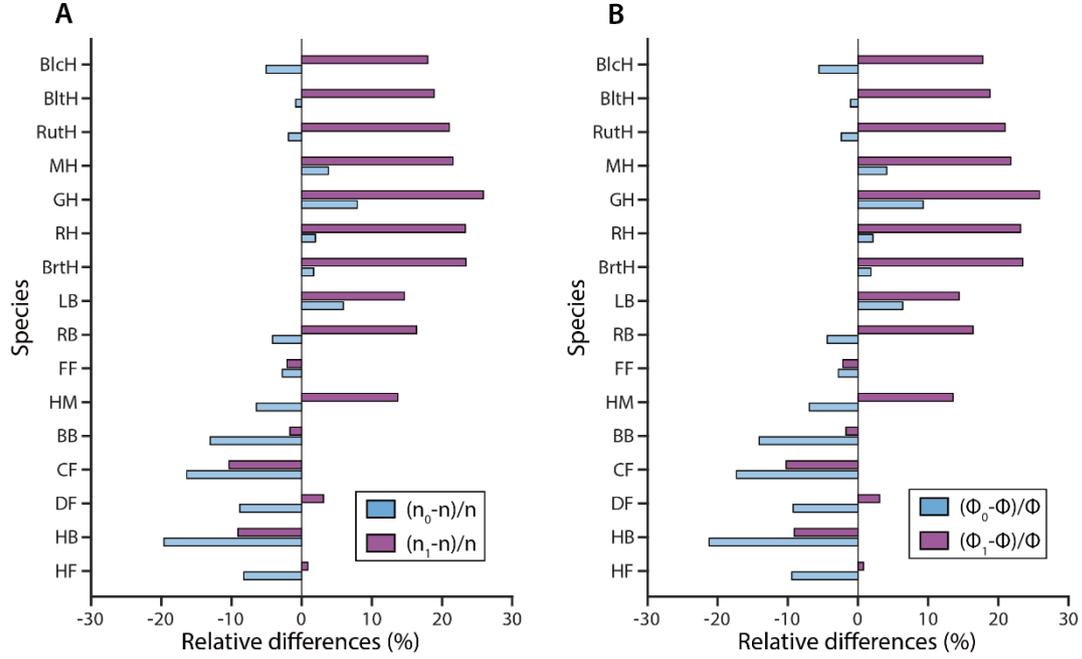

**Fig. 2. Relative differences between the theoretically optimal kinematic parameters and the measured parameters.**
**(A)** The relative differences between the optimal flapping frequencies $n_0$ (which minimize hovering power without elastic energy storage) and $n_1$ (which minimize power with complete elastic energy storage), compared to the measured flapping frequency ($n$) under the condition of a fixed measured flapping amplitude ($\Phi$). In the figure, the vertical axis represents different species, listed from top to bottom as follows: BlcH (Black-chinned hummingbird), BltH (Blue-throated hummingbird), RutH (Ruby-throated hummingbird), MH (Magnificent hummingbird), GH (Giant hummingbird), RH (Rufous hummingbird), BrtH (Broad-tailed hummingbird), LB (Ladybird), RB (Rhinoceros beetle), FF (Fruit fly), HM (Hawkmoth), BB (Bumblebee), CF (Crane fly), DF (Drone fly), HB (Honeybee), and HF (Hover fly).
**(B)** The relative differences between the optimal flapping amplitudes $\Phi_0$ (which minimize hovering power without elastic energy storage) and $\Phi_1$ (which minimize power with complete elastic energy storage) and the measured flapping amplitude ($\Phi$) under the condition of a fixed measured flapping frequency ($n$). The vertical coordinates in this figure correspond to those in (A).

As shown in Fig. 2, the measured wing kinematic parameters ($n$, $\Phi$) of hummingbirds, ladybirds (LB) and rhinoceros beetle (RB) are closer to the optimal values ($n_0$, $\Phi_0$) calculated under the condition without elastic energy storage. This suggests that during hovering flight, these species may prioritize minimizing total mechanical power—a combination of aerodynamic and inertial power—and that their capability for elastic energy storage is likely very weak. This conclusion aligns with previous research: Agrawal et al. (*27*) demonstrated that hummingbird have inferior elastic energy storage capability compared to locusts and bumble bees; Phan and Park (*14*) demonstrated that rhinoceros beetles employ a high angle of attack with low aerodynamic efficiency during hovering to reduce inertial power. Although direct research on elastic energy storage in ladybirds is currently lacking, considering that both ladybirds and rhinoceros beetles belong to the order *Coleoptera*, they may exhibit similarly limited elastic energy storage capabilities.

Unlike hummingbirds, ladybirds and rhinoceros beetles, the measured wing kinematic parameters ($n$, $\Phi$) of most insects—such as drone flies (DF), hover flies (HF), crane flies (CF), bumblebees (BB), and honeybees (HB)—are closer to the optimal values ($n_1$, $\Phi_1$) calculated under condition of complete elastic energy storage, as shown in Fig. 2. This suggests that these insects tend to minimize aerodynamic power during hovering, and likely possess strong elastic energy storage capabilities. This conclusion align with existing studies: Berman and Wang (*28*) observed that the flapping frequency of bumblebees during hovering approaches theoretical values that minimizing aerodynamic power; Wu and Sun (*40*) confirmed that drone fly wing kinematics correspond to minimum aerodynamic power states; and Corban et al. (*41*) similarly reported that honeybees adopt wing kinematics with high aerodynamic efficiency during hovering flight.

In addition, Fig. 2 illustrates that the elastic energy storage capabilities of hawkmoths and fruit flies may differ from those of the two previously mentioned groups. The measured flapping frequency ($n$) of the hawkmoth lies between $n_0$ and $n_1$, and the measured flapping amplitude ($\Phi$) lies between $\Phi_0$ and $\Phi_1$, indicating that hawkmoths may utilize only a portion of the excess kinetic energy during wing deceleration, indicating a moderate elastic energy storage capability. Supporting evidence includes Zheng et al. (*42*), who reported that hawkmoths do not operate at minimum aerodynamic power during hovering flight, while Gau et al. (*26*) and Reid et al. (*25*) independently confirmed energy recovery capability in the scutum region and wings of the hawkmoth, respectively. Fruit flies represent an exceptional case, as their measured parameters ($n$, $\Phi$) nearly coincide with both theoretical optima ($n_0$, $n_1$, $\Phi_0$, $\Phi_1$). This implies that, regardless of elastic energy storage capability, fruit flies can hover at a state close to the theoretical minimum mechanical power. The key factor contributing to this phenomenon is the fruit flies' exceptionally low wing mass (approximately 0.24% of its body weight (*43*)), which results in inertial power requirements substantially lower than aerodynamic power (*9*, *44*). Consequently, even if fruit flies possess complete elastic energy storage capability, the reduction in hovering power requirements would not exceed 2% (*9*, *10*, *45*), having a minimal impact on overall energy expenditure.

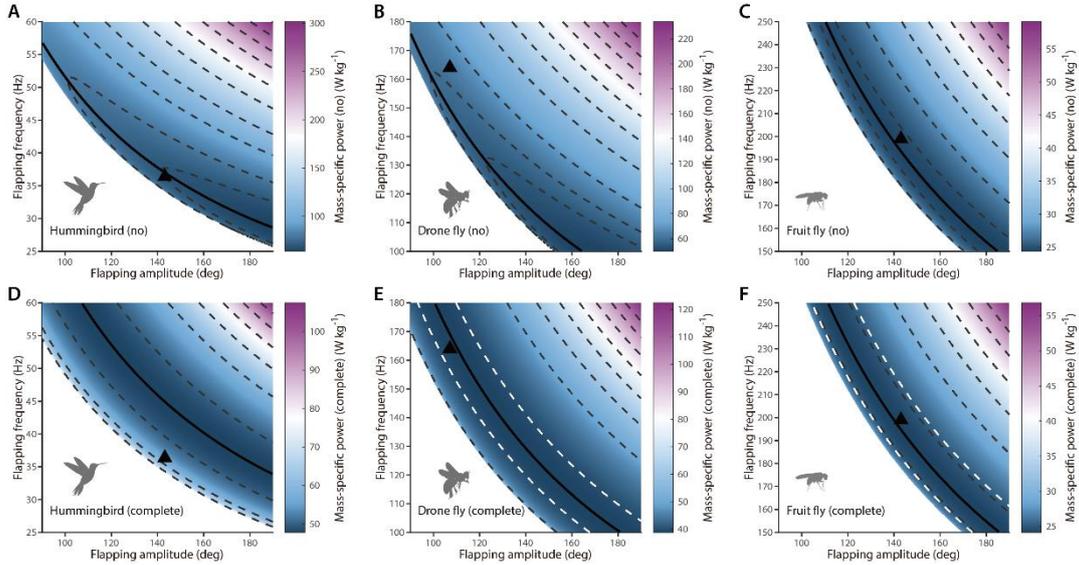

**Fig. 3. The relationship between mechanical power requirements for hovering flight and wing kinematic parameters: flapping amplitude ($\Phi$) and frequency ($n$).**
**(A-C)** Mass-specific mechanical power (mechanical power/body mass) without elastic energy storage for (A) drone fly, (B) broad-tailed hummingbird, and (C) fruit fly. Black triangles represent measured kinematic parameters ($\Phi$, $n$). Solid black curves depict the "optimal wing kinematics curve" under fixed frequency (or amplitude) constraints: for a given $n$ (or $\Phi$), the curve show the corresponding $\Phi$ (or $n$) that minimizes mechanical power without elastic energy storage. White regions indicate kinematic combinations that cannot generate sufficient lift for hovering.
**(D-F)** Mass-specific mechanical power (mechanical power/body mass) with complete elastic energy storage for (D) drone fly, (E) broad-tailed hummingbird, and (F) fruit fly. Black triangles indicate measured kinematic parameters ($\Phi$, $n$). Solid black curves represent the "optimal wing kinematics curve" under fixed frequency (or amplitude) constraints: for a given $n$ (or $\Phi$), the curve shows the corresponding $\Phi$ (or $n$) that minimizes mechanical power with complete elastic storage. The white dashed line represents the contour corresponding to 1.05 times the minimum mass-specific mechanical power. White regions indicate kinematic combinations that cannot generate sufficient lift for hovering.

To visualize these results, Fig. 3 illustrates the relationships among flapping amplitude ($\Phi$), flapping frequency ($n$), and mechanical power requirements for hovering under two extreme elastic storage conditions for the broad-tailed hummingbird (representing weak elastic storage), the drone fly (representing strong elastic storage), and the fruit fly (an exceptional case). In Fig. 3, the solid black curves denote the "optimal wing kinematics curve", which minimizes mechanical power under fixed flapping frequency (or flapping amplitude) constraints for each elastic storage scenario. Black triangles indicate the measured kinematics and their corresponding mass-specific mechanical power. Fig. 3A and 3D clearly demonstrate that the measured data point for the broad-tailed hummingbird is close to the optimal curve without elastic storage but distant from the optimal curve with complete elastic storage. Conversely, the measured data of the drone fly exhibit the opposite pattern, as shown in Fig. 3B and E. This suggests that if their actual storage capability contradicted our inferences, they would hover in states of high energy expenditure, significantly deviating from the theoretical optima. Meanwhile, the fruit fly's measured data point approximates both optimal curves simultaneously (Fig. 3C and F), confirming that—regardless of whether it possesses elastic storage capability—its adopted wing kinematics ensure hovering with high energy efficiency.

All the aforementioned analyses of elastic energy storage capability have been limited to specific hovering equilibrium conditions. However, when these conditions change一such as increased body weight after feeding, carrying loads, or migrating to high-altitude environments with low air density—small flying animals typically adjust wing kinematic parameters (usually by modifying flapping amplitude, flapping frequency, or angle of attack) to maintain hovering. This scenario provides a novel perspective for evaluating their elastic storage capability. We hypothesize that: (i) the newly selected kinematics ($n_{new}$, $\Phi_{new}$) will align with the "optimal wing kinematics curve" for the new equilibrium condition, and (ii) the type of optimal curve remains determined by the inherent elastic storage capability of small flying animals—species with weak capability correspond to the minimum total mechanical power curve, whereas species with capability align with the minimum aerodynamic power curve. Notably, even under these hypothetical conditions, small flying animals may employ distinct kinematic adjustment strategies (e.g., increasing $\Phi$ alone, $n$ alone, or both). Crucially, because aerodynamic and inertial power exhibit differential sensitivity to wing kinematic, each strategy incurs distinct mechanical power requirements. To conserve energy, small flyers are expected to adopt the adjustment strategy with lower energy expenditure, and this strategy should correlate with their elastic energy storage capability. We propose leveraging this relationship to inversely infer their elastic storage capability.

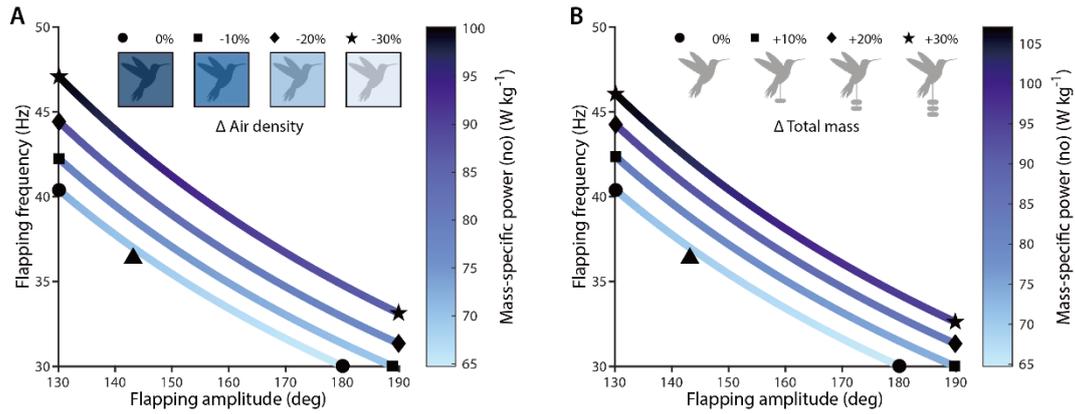

**Fig. 4. Variation of the "optimal wing kinematics curve" of hummingbird as hovering equilibrium condition changes.**

**(A)** Variation of the "optimal wing kinematics curve" without elastic energy storage as air density (*ρ*) decreases. The curves with endpoints shaped as circles, squares, diamonds, and pentagrams correspond to 0%, 10%, 20%, and 30% reduction in air density, respectively. The black triangle indicates the measured $\Phi$ and $n$ of the hummingbird when the hovering equilibrium condition remains unchanged.

**(B)** Variation of the "optimal wing kinematics curve" without elastic storage as total mass (body mass + payload) increases. The curves with endpoints shaped as circles, squares, diamonds, and pentagrams correspond to 0%, 10%, 20%, and 30% increases in total mass (i.e., payload). The black triangle indicates the measured $\Phi$ and $n$ of hummingbird when the hovering equilibrium condition remains unchanged.

To validate this hypothesis, we examined the relationship between wing kinematic adjustment strategies and mechanical power requirement during hovering under conditions of reduced air density and increased load mass (i.e., total mass). We selected the broad-tailed hummingbird (representing weak elastic energy storage) and the drone fly (representing strong elastic energy storage) as representative species, as shown in Fig. 4 and 5. Fig. 4 presents the "optimal wing kinematics curve" for various hovering states in broad-tailed hummingbirds assuming no elastic energy storage, while Fig. 5 displays the corresponding optimal curve for the drone fly, assuming complete elastic energy storage.

Fig. 4 demonstrates that, assuming no elastic energy storage, a decrease in air density or an increase in load mass prompts an adjustment strategy involving solely increasing the flapping amplitude (*Φ*). This strategy corresponds to lower mechanical power requirements. In contrast, simultaneously increasing both flapping amplitude and frequency, or increasing frequency alone, results in significantly higher mechanical power requirement. These findings align closely with the actual behavior of hummingbirds. When faced with reduced air density or light loads, hummingbirds substantially increase flapping amplitude while maintaining nearly constant frequency (*46–50*).

Only when the load approaches or exceeds their body mass do they significantly increase frequency. At this point, the hummingbirds' wing tips at the extreme position of the downstroke touch or even cross each other (*37*). This behavioral pattern indicates that hummingbirds tend to adopt the "increase $\Phi$ only" strategy to minimize mechanical power expenditure, resorting to frequency adjustments only when necessary. These observations provide compelling evidence that hummingbirds have very limited elastic energy storage capability.

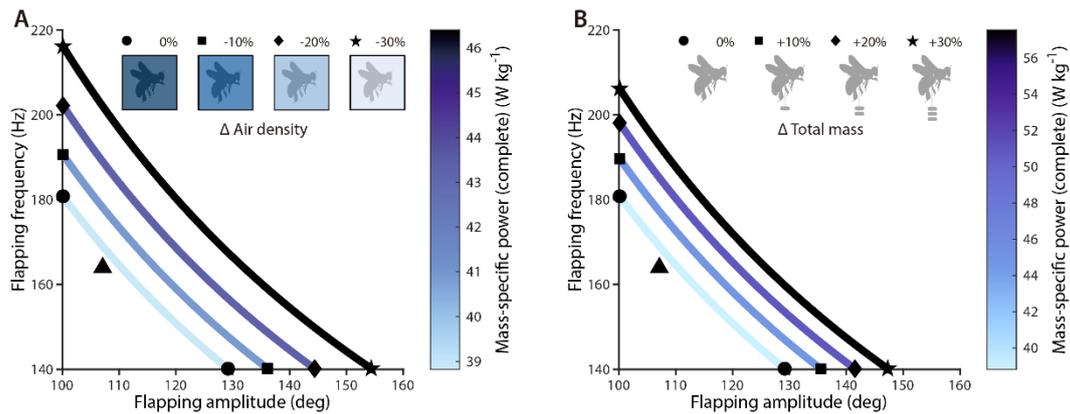

**Fig. 5. Variation of the "optimal wing kinematics curve" of drone fly as hovering equilibrium condition changes.**

**(A)** Variation of the "optimal wing kinematics curve" with complete elastic energy storage as air density ($\rho$) decreases. The curves with endpoints shaped as circles, squares, diamonds, and pentagrams correspond to 0%, 10%, 20%, and 30% reduction in air density, respectively. Black triangle indicates the measured $\Phi$ and $n$ of the drone fly when the hovering equilibrium condition remains unchanged.

**(B)** Variation of the "optimal wing kinematics curve" with complete elastic storage as total mass (body mass + payload) increases. The curves with endpoints shaped as circles, squares, diamonds, and pentagrams correspond to 0%, 10%, 20%, and 30% increases in total mass (i.e., payload). Black triangle indicates the measured $\Phi$ and $n$ of drone fly when the hovering equilibrium condition remains unchanged.

Fig. 5 demonstrates that for species with strong elastic energy storage capabilities, various wing kinematic adjustment strategies result in only slight differences in mechanical power (i.e., aerodynamic power) requirements. Consequently, these species exhibit considerable flexibility in selecting adjustment strategies if energy minimization is the sole consideration. However, experimental observations reveal striking interspecies differences in kinematic adjustment patterns. Bumblebees predominantly increase flapping amplitude when facing reduced air density conditions (*51*). Under load-carrying scenarios, bumblebees' flapping amplitude shows exhibits a positive correlation with payload mass, while flapping frequency displays irregular fluctuations or even decreases (*52*). This flexibility in wing kinematic adjustment aligns with our predictions for species with strong elastic energy storage capability. In contrast, honeybees display different response patterns. Under decreased air density conditions, they exclusively increase flapping amplitude while maintaining an essentially constant flapping frequency (*53*, *54*). More notably, even when subjected to light loads (18% of body mass), neither $\Phi$ nor $n$ demonstrates significant changes (*55*). These behaviors make reliable assessment of their energy storage capability particularly challenging using this method. Furthermore, the absence of systematic kinematic studies on hoverflies, craneflies, and hawkmoths prevents evaluation of their elastic storage capabilities through this approach.

## DISCUSSION

Based on the assumption that hovering animals tend to adopt kinematics that minimize energy expenditure, our results not only enable the assessment of elastic energy storage capabilities across species but also identify the necessary parameters and conditions for determining the presence of elastic energy storage. Flapping amplitude ($\Phi$) emerges as a critical indicator of elastic storage capability. Species with weak elastic storage capability consistently exhibit a strong preference for large flapping amplitudes ($\Phi > 140°$). As illustrated in Fig. 3A, along the "optimal wing kinematics curve" under conditions with no elastic storage, mechanical power requirements progressively decrease with increasing $\Phi$. Therefore, when physiological structures allow, species with weak elastic energy storage capability adopt large flapping amplitudes to minimize energy expenditure— a strategy that aligns precisely aligned with experimental observations in hummingbirds, ladybirds and rhinoceros beetles (Table S1).

In contrast, species with strong elastic storage capabilities consistently prefer smaller flapping amplitudes. As illustrated in Fig. 3E, mechanical power requirements under complete elastic energy storage conditions—equivalent to aerodynamic power—show negligible variation along the "optimal wing kinematics curve" across different $\Phi$ values. Furthermore, even moderate deviations from this optimal curve result in only slight increases in mechanical power. These characteristics collectively indicate that species with strong elastic energy storage capabilities possess substantial flexibility in selecting the combinations of $\Phi$ and $n$ within a broad operational range (demarcated by white dashed lines in Fig. 3E). This characteristic enables them to adopt either large-amplitude and low-frequency, or small-amplitude and high-frequency hovering strategies.

However, experimental evidence reveals that insects predominantly select the latter (low-amplitude and high-frequency), which we attribute to two possible constraints. First, physiological limitations restrict maximum flapping amplitudes in some insects compared to hummingbirds, ladybirds or rhinoceros beetles. For instance, carpenter bees reach only about 140° (*18*), and bumblebees about 145°–150° (*51*). Second, most insects are unlikely to to significantly adjust their flapping frequency. Apart from the honeybees mentioned above, carpenter bees (*18*) and orchid bees (*56*) maintain nearly constant flapping frequencies when air density changes, and fruit flies (*31*, *45*) preserve consistent flapping frequency during hovering and ascending flight. Consequently, large-amplitude strategies would severely constrain aerodynamic force modulation capability in these insects, making the small-amplitude, high-frequency approach operationally advantageous.

As mentioned earlier, hummingbirds maintain hovering when facing increased loading or reduced air density primarily through substantial elevation of flapping amplitude ($\Phi$) while maintaining nearly constant flapping frequency ($n$)—an energy-saving strategy consistent with their limited elastic storage capability. Although certain insects like honey bees exhibit similar adjustment strategy under reduced air density, we believe a fundamental distinction between their behaviors. Crucially, despite hummingbirds exhibiting relatively constant flapping frequencies (*57*), extensive studies confirm their capability for active, substantial frequency modulation when encountering environmental variations (e.g., temperature or weather changes) (*36*, *46*, *58*), extreme loading conditions (*37*, *49*, *59*), or during maneuvering flight (*60*). This demonstrates that hummingbirds' preferential adoption of the "increase $\Phi$ only" strategy represents an energetically optimal choice reflecting their inherently limited elastic storage capability, rather than resulting from physiological constraints on frequency adjustment.

In contrast, the resonant flapping mechanism inherent to most insects—where thoracic oscillations drive wing motion through structural resonance (*61*, *62*)—severely limits sustained and substantial adjustments in flapping frequency. Consequently, even species with complete elastic energy storage capabilities exhibit limited flexibility in kinematic adjustments, particularly regarding frequency modulation, due to these physiological constraints.

Furthermore, according to the aforementioned research, if hummingbirds indeed have a relatively weak elastic energy storage, they must prioritize the low energy expenditure strategy of "increase $\Phi$ only" which has been strongly supported by previous experimental studies. Altshuler et al. (49) demonstrated that hummingbirds showed no significant increase in flapping frequency even when air density dropped below sustainable hovering thresholds. In contrast, under normal air density with extreme loading, they simultaneously increased both flapping amplitude and frequency to generate sufficient lift. Notably, this extreme-load hovering could be maintained for only 1–2 seconds, followed by immediate panting or noticeably heavier breathing, suggesting that the energy demands likely exceeded aerobic metabolic limits. This compelling comparison indicates that hummingbirds preferentially avoid the energy-intensive strategy of increasing flapping frequency even under extreme conditions (critically low air densities that preclude hovering), resorting to simultaneous enhancement of both amplitude and frequency only when compelled by extreme loading.

This study not only introduces a novel approach for assessing elastic energy storage capability but also provides valuable insights for the design of flapping wing micro air vehicles (FWMAVs). First, the simplified power model enables rapid evaluation of the mechanical power requirements for FWMAVs. For instance, the Nano Hummingbird (Saturn prototype) (63)—a FWMAV without elastic energy storage—is estimated by the model to consume 1.48 W of mechanical power during hovering (mechanical power loading of 11.81 gf $W^{-1}$), whereas actual measurements indicate 1.86 W (mechanical power loading of 9.40 gf $W^{-1}$) (see Supplementary Materials). This discrepancy primarily arises because the model does not account for the power consumption of the flapping mechanism. Nevertheless, the model provides a theoretical reference close to the minimum power requirement for FWMAVs, with the difference from actual values essentially reflecting the additional power consumption of the drive structure. Therefore, this model and its evaluation results offer important design guidance for the energy and power systems of FWMAVs, such as battery and motor selection.

This model can also assist in optimizing the wing kinematic parameters of FWMAVs to achieve hovering flight with low power requirement and long endurance. Currently, most FWMAVs operate with fixed flapping amplitudes. Using our model, we can determine the optimal flapping frequency that minimizes hovering power requirements. For example, considering the Nano Hummingbird (Saturn prototype), when maintaining its flapping amplitude of 180°, the model predicts an optimal flapping frequency of 28.2 Hz, which closely matches the actual operational frequency of 27.5 Hz. Furthermore, Fig. 3B demonstrates that, in the absence of elastic energy storage, increasing the flapping amplitude effectively reduces mechanical power requirements. This finding validates the design rationale behind the large flapping amplitudes adopted by most hummingbird-inspired FWMAVs, including COLIBRI (*64*), KUBeetle (*65*), and Robotic Hummingbird (*66*).

Moreover, the model enables the estimation of excess kinetic energy during wing deceleration and its proportion relative to the total mechanical power requirement, which determines the theoretical maximum benefit achievable through elastic energy storage design. For the Nano Hummingbird (Saturn prototype), the dissipated excess kinetic energy accounts for only about 18.1% of the total mechanical work during hovering. Considering that implementing elastic structures would increase the vehicle's mass and complexity while reducing its reliability—along with the fatigue challenges posed by high-frequency cyclic deformation of elastic elements—the practical benefits for FWMAV may be limited. This likely explains why a few hummingbird-inspired FWMAVs incorporating elastic storage (*67*, *68*) have failed to demonstrate significant improvements in power consumption or flight endurance. However, the advantages of elastic storage have been validated in smaller-scale or lower-lift/thrust flapping prototypes. For instance, Lau et al. (*69*) developed a flexible mechanism operating at 15 Hz and generating 0.5 gf of thrust, achieving up to a 31% reduction in mechanical power compared to flapping systems without elastic storage. These findings suggest that the application of elastic storage in FWMAVs requires thorough evaluation based on specific design objectives, where trade-offs between power savings and system penalties must be carefully balanced.

It is important to clarify that the simplified power model developed in this study is based on the following key assumptions: (i) the animals achieve hovering solely through a single pair of wings; (ii) the flapping amplitude of the animals is sufficiently large (> 90°); and (iii) the stroke plane is approximately horizontal. These fundamental premises inherently restrict the model's applicability to certain biological systems. Specifically, the model is not suitable for species that use two independent pairs of wings, such as dragonflies (*70*), butterflies (*71*), and locusts (*72*); species with very small flapping amplitudes, such as mosquitoes (*73*, *74*); species exhibiting steep stroke planes, such as long-tongued bats (*75*); or very small insects without defined stroke planes, such as tiny wasps (*76*). This limitation highlights the need for future research to expand the model's framework by incorporating a broader range of small flying animals, particularly those with unconventional kinematic patterns, to improve our comprehensive understanding of biological flight energetics and the design of (FWMAVs).

## MATERIALS AND METHODS

To facilitate the calculation of optimal kinematic parameters ($n_0$, $n_1$, $\Phi_0$, $\Phi_1$), we developed a simplified model to rapidly assess the mechanical power requirements during hovering, as detailed below.

**Simplification of flapping motion**

The flapping motion of the wings of small flying animals during hovering can be simplified as a combination of translational and rotational motions. The flapping angle ($\phi$) follows an approximately sinusoidal trajectory over one flapping cycle (*33*). Defining each cycle as beginning with the upstroke, $\phi(t)$ can be expressed as:

When a small flying animal hovers, the flapping motion of its wings can be simplified as a combination of translation and rotational movements. The flapping angle ($\phi$) varies with time ($t$) during one flapping cycle, approximately following a sinusoidal curve (*33*). If each flapping cycle is defined as starting with the downstroke, then $\phi(t)$ can be expressed as:

$$\phi(t) = -\frac{\Phi}{2}\cos(2\pi n t) \tag{1}$$

where, $\Phi(=\phi_{\max} - \phi_{\min})$ represents the flapping amplitude, and $n$ is the flapping frequency. From Eq. 1, the flapping angular velocity ($\dot\phi$) and angular acceleration ($\ddot\phi$) can be obtained, as shown in Eq. 2 and 3.

$$\dot\phi(t) = \Phi\pi n \sin(2\pi n t) \tag{2}$$

$$\ddot\phi(t) = 2\Phi\pi^2 n^2 \cos(2\pi n t) \tag{3}$$

**Simplification of Instantaneous Mechanical Power**

The instantaneous mechanical power ($P_\mathrm{m}$) required for a single wing to flap consists of the instantaneous inertial power ($P_i$) and aerodynamic power ($P_a$):

$$P_m(t) = P_i(t) + P_a(t) \tag{4}$$

$P_i$ consists of two components: $P_{i,tr}$ (the inertial power associated with translational motion) and $P_{i,ro}$ (the inertial power associated with rotational motion). Since the moment of inertia of the wing about the rotational axis is significantly smaller than that about the wing root, $P_{i,ro}$ can be neglected (*10*). Therefore, $P_i$ can be expressed as:

$$P_i(t) \approx P_{i,tr}(t) = I\ddot{\phi}(t)\dot{\phi}(t) = 2\pi^3 I \Phi^2 n^3 \cos(2\pi nt)\sin(2\pi nt) \tag{5}$$

In Eq. 5, $I$ is the moment of inertia of the wing about the flapping axis.

Similarly, $P_a$ consists of two parts: $P_{a,tr}$ (the aerodynamic power associated with the translational motion) and $P_{a,ro}$ (the aerodynamic power associated with the rotational motion). When the flapping amplitude is large ($\Phi > 90°$), compared to $P_{a,tr}$, $P_{a,ro}$ is smaller and can therefore be neglected (*10*). Additionally, during the translational motion, the aerodynamic center of the wing is approximately located at the position of the second moment of area $R_{2,S}(=\hat{r}_{2,S}R)$ along the spanwise direction, where $R$ is the wing length and $\hat{r}_{2,S}$ is the nondimensional radius of the second moment of wing area. Consequently, the instantaneous aerodynamic power $P_a(t)$ can be further expressed as:

$$P_a(t) \approx P_{a,tr}(t) = D_{tr}(t) R_{2,S} |\dot{\phi}(t)| \tag{6}$$

where $D_{tr}$ represents the instantaneous aerodynamic drag experienced by a single wing during translational motion, and it can be further expressed as the product of the dimensionless instantaneous translational drag coefficient ($C_{D,tr}$) and reference quantities:

$$D_{tr}(t) = C_{D,tr}(t) \frac{1}{2} \rho \bar{U}_{ref}^2 S \propto C_{D,tr}(t) \bar{U}_{ref}^2 \tag{7}$$

where $\rho$ represents the air density, $S$ is the area of a single wing, and $\bar{U}_{ref} (= 2\Phi n R_{2,S})$ is the reference velocity.

Furthermore, in the quasi-steady model of flapping wings, it is generally accepted that $D_{tr}$ is positively correlated with the square of the instantaneous velocity $U_{R_{2,S}}(= R_{2,S}\dot{\phi}(t))$ at $R_{2,S}$, that is:

$$D_{\text{tr}}(t) \propto U_{R_{2,S}}^2(t) = (R_{2,S}\dot{\phi}(t))^2 = \bar{U}_{\text{ref}}^2 \sin^2(2\pi nt) \tag{8}$$

Therefore, from Eq. 7 and 8, it can be concluded that $C_{D,\text{tr}}$ satisfies the following relationship.

$$C_{D,\text{tr}}(t) \propto \sin^2(2\pi nt) = \frac{1 - \cos(4\pi nt)}{2} \tag{9}$$

We further consider that: (i) the stroke planes of most small flying animals are nearly horizontal, and the mean lift and drag generated during the upstroke and downstroke are approximately equal; (ii) compared with translational motion, the drag generated during rotational motion can be neglected; and, (iii) the mean value on the right side of Eq. 9 over one flapping cycle ($1/n$) is $1/2$. Based on these three points, we can further express $C_{D,\text{tr}}$ as:

$$C_{D,\text{tr}} \approx \bar{C}_D[1 - \cos(4\pi nt)] \tag{10}$$

where $\bar{C}_D$ represents the mean drag coefficient. Combining Eq. 6, 7, and 10, the instantaneous aerodynamic power ($P_a$) required for the single wing flapping can be expressed as:

$$P_a(t) = 4\pi\rho\Phi^3 n^3 R_{2,S}^3 S\bar{C}_D |\sin^3(2\pi nt)| \tag{11}$$

So far, we have derived the expressions for $P_i(t)$ and $P_a(t)$.

**Calculation of Mean Mechanical Power Requirement**

By substituting $P_i(t)$ and $P_a(t)$ in Eq. 4 with Eq. 5 and 11, respectively, the instantaneous mechanical power requirement ($P_m(t)$) for a (single) wing can be expressed as:

$$P_m(t) = A|\sin^3(\omega t)| + B\cos(\omega t)\sin(\omega t) \tag{12}$$

where $A$ ($= 4\pi\rho\Phi^3 n^3 R_{2,S}^3 S\bar{C}_D$) and $B$ ($= 2\pi^3 I\Phi^2 n^3$) are defined as the aerodynamic power parameter and the inertial power parameter, respectively, and $\omega$ ($= 2\pi n$) is the angular frequency.

With the expression for $P_m$ obtained, we can further calculate the mean mechanical power requirements under two extreme elastic energy storage conditions. For the no elastic energy storage scenario, the mean mechanical power requirement ($\bar{P}_{m,0}$) corresponds to the mean value of the positive components of $P_m$ (i.e., $P_m^+$, blue area in Fig. 1B) over one flapping cycle (Eq. 13). In contrast, under complete elastic energy storage conditions, the mean mechanical power requirement ($\bar{P}_{m,1}$), equivalent to the mean aerodynamic power ($\bar{P}_a$), equals the mean value of $P_m$ (or, $P_a$) over one flapping cycle (Eq. 14).

$$\bar{P}_{m,0} = n \int_0^{1/n} P_m^+(t)dt = \frac{1}{\pi}\left[\frac{2A}{3} - \frac{B^3}{12A^2} + \frac{(B^2 + 4A^2)^{\frac{3}{2}}}{12A^2}\right] \quad (13)$$

$$\bar{P}_{m,1} = n \int_0^{1/n} P_m(t)dt = n \int_0^{1/n} P_a(t)dt = \frac{4A}{3\pi} \quad (14)$$

Additionally, we quantified the proportion of excess kinetic energy (represented by the purple area in Fig. 1B) to the total mechanical power requirement under condition without elastic storage. This proportion was calculated as $(\bar{P}_{m,0} - \bar{P}_{m,1})/\bar{P}_{m,0}$.

**Determination of the Mean Drag Coefficient**

After obtaining the morphological and kinematic parameters of small flying animals, it is essential to first determine the mean drag coefficient during the flapping process, $\bar{C}_D$, before estimating the mechanical power requirements for hovering flight according to Eqs. (13) and (14). The procedure for calculating $\bar{C}_D$ is as follows. First, under hovering conditions, the mean lift ($\bar{L}$) must equal the weight of the animal, from which the mean lift coefficient ($\bar{C}_L$) can be calculated (Eq. (15)). Then, using the previously obtained $\bar{C}_L$-$\bar{\alpha}$ curve (where $\bar{\alpha}$ is the mean angle of attack), $\bar{\alpha}$ is determined. Subsequently, the mean drag coefficient $\bar{C}_D$ is found from the previously obtained $\bar{C}_D$-$\bar{\alpha}$ curve. Both curves were derived via CFD methods, with detailed information provided in the Supplementary Materials.

$$\bar{C}_L = \frac{2\bar{L}}{\rho \bar{U}_{\text{ref}}^2 (2S)} \quad (15)$$

During the calculation of $\bar{C}_L$ and $\bar{C}_D$, we treated these coefficients as functions solely of the Reynolds number, independent of wing geometry. This simplification is justified by previous studies demonstrating that when using the average velocity at the second moment of wing area, $U_{R_{2,S}}$, as the reference velocity, both $\bar{C}_L$ and $\bar{C}_D$ remain remarkably insensitive to wing geometry across the characteristic aspect ratio range (2.8-5.5) typical of most small flying animals (*77*). Furthermore, wing geometry exerts minimal influence on the lift-to-drag ratio ($\bar{C}_L/\bar{C}_D$) in this regime (*78*).

# REFERENCES

1. T. Weis-Fogh, Energetics of Hovering Flight in Hummingbirds and in Drosophila. *Journal of Experimental Biology* **56**, 79–104 (1972).

2. M. H. Dickinson, M. S. Tu, The function of dipteran flight muscle. *Comparative Biochemistry and Physiology Part A: Physiology* **116**, 223–238 (1997).

3. D. Labonte, N. C. Holt, Elastic energy storage and the efficiency of movement. *Current Biology* **32**, R661–R666 (2022).

4. F. G. Stiles, Time, Energy, and Territoriality of the Anna Hummingbird ( *Calypte anna* ). *Science* **173**, 818–821 (1971).

5. J.-R. Martin, A portrait of locomotor behaviour in Drosophila determined by a video-tracking paradigm. *Behavioural Processes* **67**, 207–219 (2004).

6. R. Dudley, C. P. Ellington, Mechanics of forward flight in bumblebees: II. Quasi-steady lift and power requirements. *Journal of Experimental Biology* **148**, 53–88 (1990).

7. J. Wu, M. Sun, Unsteady aerodynamic forces and power requirements of a bumblebee in forward flight. *ACTA MECH SINICA* **21**, 207–217 (2005).

8. C. J. Clark, R. Dudley, Hovering and Forward Flight Energetics in Anna's and Allen's Hummingbirds. *Physiological and Biochemical Zoology* **83**, 654–662 (2010).

9. H. J. Zhu, M. Sun, Kinematics measurement and power requirements of fruitflies at various flight speeds. *Energies* **13**, 4271 (2020).

10. S. Mao, D. Gang, Lift and power requirements of hovering insect flight. *Acta Mechanica Sinica* **19**, 458–469 (2003).

11. Y. Z. Lyu, M. Sun, Power requirements for the hovering flight of insects with different sizes. *Journal of Insect Physiology* **134**, 104293 (2021).

12. C. P. Ellington, The aerodynamics of hovering insect flight. VI. Lift and power requirements. *Philosophical Transactions of the Royal Society of London. B, Biological Sciences* **305**, 145–181 (1984).

13. M. H. Dickinson, J. R. Lighton, Muscle efficiency and elastic storage in the flight motor of Drosophila. *Science* **268**, 87–90 (1995).

14. H. V. Phan, H. C. Park, Wing inertia as a cause of aerodynamically uneconomical flight with high angles-of-attack in hovering insects. *Journal of Experimental Biology*, jeb.187369 (2018).

15. T. Weis-Fogh, Quick Estimates of Flight Fitness in Hovering Animals, Including Novel Mechanisms for Lift Production. *Journal of Experimental Biology* **59**, 169–230 (1973).

16. C. P. Ellington, Power and efficiency of insect flight muscle. *Journal of Experimental Biology* **115**, 293–304 (1985).

17. D. J. Wells, Muscle Performance in Hovering Hummingbirds. *Journal of Experimental Biology* **178**, 39–57 (1993).

18. S. P. Roberts, J. F. Harrison, R. Dudley, Allometry of kinematics and energetics in carpenter bees (Xylocopa varipuncta) hovering in variable-density gases. *Journal of Experimental Biology* **207**, 993–1004 (2004).

19. J. Song, H. Luo, T. L. Hedrick, Three-dimensional flow and lift characteristics of a hovering ruby-throated hummingbird. *J. R. Soc. Interface.* **11**, 20140541 (2014).

20. Y. Liu, M. Sun, Wing kinematics measurement and aerodynamics of hovering dronefies. *Journal of Experimental Biology* **211**, 2014–2025 (2008).

21. T. Weis-Fogh, R. M. Alexander, T. J. Pedley, The sustained power output from striated muscle. *Scale effects in animal locomotion*, 511–525 (1977).

22. C. J. Pennycuick, M. A. Rezende, The Specific Power Output Of Aerobic Muscle, Related To The Power Density Of Mitochondria. *Journal of Experimental Biology* **108**, 377–392 (1984).

23. R. K. Josephson, Contraction Dynamics and Power Output of Skeletal Muscle. *Annu. Rev. Physiol.* **55**, 527–546 (1993).

24. F.-O. Lehmann, The constraints of body size on aerodynamics and energetics in flying fruit flies: an integrative view. *Zoology* **105**, 287–295 (2002).

25. H. E. Reid, R. K. Schwab, M. Maxcer, R. K. D. Peterson, E. L. Johnson, M. Jankauski, Wing flexibility reduces the energetic requirements of insect flight. *Bioinspir. Biomim.* **14**, 056007 (2019).

26. J. Gau, N. Gravish, S. Sponberg, Indirect actuation reduces flight power requirements in Manduca sexta via elastic energy exchange. *Journal of the Royal Society Interface* **16**, 20190543 (2019).

27. S. Agrawal, B. W. Tobalske, Z. Anwar, H. Luo, T. L. Hedrick, B. Cheng, Musculoskeletal wing-actuation model of hummingbirds predicts diverse effects of primary flight muscles in hovering flight. *Proc. R. Soc. B.* **289**, 20222076 (2022).

28. G. J. Berman, Z. J. Wang, Energy-minimizing kinematics in hovering insect flight. *J. Fluid Mech.* **582**, 153–168 (2007).

29. B. Cheng, B. W. Tobalske, D. R. Powers, T. L. Hedrick, Y. Wang, S. M. Wethington, G. T.-C. Chiu, X. Deng, Flight mechanics and control of escape manoeuvres in hummingbirds II. Aerodynamic force production, flight control and performance limitations. *Journal of*

*Experimental Biology*, jeb.137570 (2016).

30. C. Shen, M. Sun, Power Requirements of Vertical Flight in the Dronefly. *J Bionic Eng* **12**, 227–237 (2015).

31. C. Shen, Y. Liu, M. Sun, Lift and power in fruitflies in vertically-ascending flight. *Bioinspir. Biomim.* (2018).

32. C. P. Ellington, The aerodynamics of hovering insect flight. II. Morphological parameters. *Philosophical Transactions of the Royal Society of London. B, Biological Sciences* **305**, 17–40 (1984).

33. C. P. Ellington, The aerodynamics of hovering insect flight. III. Kinematics. *Philosophical Transactions of the Royal Society of London. B, Biological Sciences* **305**, 41–78 (1984).

34. M. J. Fernandez, *Flight Performance and Comparative Energetics of the Giant Andean Hummingbird (Patagona Gigas)* (University of California, Berkeley, 2010).

35. M. N. Haque, B. Cheng, B. W. Tobalske, H. Luo, Hummingbirds use wing inertial effects to improve manoeuvrability. *Journal of The Royal Society Interface* **20**, 20230229 (2023).

36. P. Chai, A. C. Chang, R. Dudley, Flight Thermogenesis and Energy Conservation in Hovering Hummingbirds. *Journal of Experimental Biology* **201**, 963–968 (1998).

37. P. Chai, D. Millard, Flight and size constraints: hovering performance of large hummingbirds under maximal loading. *Journal of Experimental Biology* **200**, 2757–2763 (1997).

38. A. R. Ennos, The kinematics and aerodynamics of the free flight of some Diptera. *Journal of Experimental Biology* **142**, 49–85 (1989).

39. S. Oh, B. Lee, H. Park, H. Choi, S.-T. Kim, A numerical and theoretical study of the aerodynamic performance of a hovering rhinoceros beetle ( *Trypoxylus dichotomus* ). *J. Fluid Mech.* **885**, A18 (2020).

40. J. H. Wu, M. Sun, Wing Kinematics in a Hovering Dronefly Minimize Power Expenditure. *Journal of Insect Science* **14** (2014).

41. B. Corban, M. Bauerheim, T. Jardin, Discovering optimal flapping wing kinematics using active deep learning. *J. Fluid Mech.* **974**, A54 (2023).

42. L. Zheng, T. L. Hedrick, R. Mittal, A multi-fidelity modelling approach for evaluation and optimization of wing stroke aerodynamics in flapping flight. *J. Fluid Mech.* **721**, 118–154 (2013).

43. S. Vogel, Flight in Drosophila: I. Flight performance of tethered flies. *Journal of Experimental Biology* **44**, 567–578 (1966).

44. M. Sun, J. Tang, Lift and power requirements of hovering flight in *Drosophila virilis*. *Journal of Experimental Biology* **205**, 2413–2427 (2002).

45. X. Meng, Y. Liu, M. Sun, Aerodynamics of Ascending Flight in Fruit Flies. *J Bionic Eng* **14**, 75–87 (2017).

46. D. J. Wells, Ecological correlates of hovering flight of hummingbirds. *Journal of Experimental Biology* **178**, 59–70 (1993).

47. P. Chai, R. Dudley, Limits to vertebrate locomotor energetics suggested by hummingbirds hovering in heliox. *Nature* **377**, 722–725 (1995).

48. P. Chai, R. Dudley, Limits to Flight Energetics of Hummingbirds Hovering in Hypodense and Hypoxic Gas Mixtures. *Journal of Experimental Biology* **199**, 2285–2295 (1996).

49. D. L. Altshuler, K. C. Welch Jr, B. H. Cho, D. B. Welch, A. F. Lin, W. B. Dickson, M. H. Dickinson, Neuromuscular control of wingbeat kinematics in Anna's hummingbirds (Calypte anna). *Journal of Experimental Biology* **213**, 2507–2514 (2010).

50. S. Mahalingam, K. C. Welch, Neuromuscular control of hovering wingbeat kinematics in response to distinct flight challenges in the ruby-throated hummingbird ( *Archilochus colubris* ). *Journal of Experimental Biology*, jeb.089383 (2013).

51. M. E. Dillon, R. Dudley, Surpassing Mt. Everest: extreme flight performance of alpine bumble-bees. *Biol. Lett.* **10**, 20130922 (2014).

52. S. A. Combes, S. F. Gagliardi, C. M. Switzer, M. E. Dillon, Kinematic flexibility allows bumblebees to increase energetic efficiency when carrying heavy loads. *Sci. Adv.* **6**, eaay3115 (2020).

53. D. L. Altshuler, W. B. Dickson, J. T. Vance, S. P. Roberts, M. H. Dickinson, Short-amplitude high-frequency wing strokes determine the aerodynamics of honeybee flight. *Proc. Natl. Acad. Sci. U.S.A.* **102**, 18213–18218 (2005).

54. J. T. Vance, D. L. Altshuler, W. B. Dickson, M. H. Dickinson, S. P. Roberts, Hovering flight in the honeybee Apis mellifera: kinematic mechanisms for varying aerodynamic forces. *Physiological and Biochemical Zoology* **87**, 870–881 (2014).

55. E. Feuerbacher, J. H. Fewell, S. P. Roberts, E. F. Smith, J. F. Harrison, Effects of load type (pollen or nectar) and load mass on hovering metabolic rate and mechanical power output in the honey bee Apis mellifera. *Journal of Experimental Biology* **206**, 1855–1865 (2003).

56. R. Dudley, Extraordinary Flight Performance of Orchid Bees (Apidae: Euglossini) Hovering in Heliox (80% He/20% O2). *Journal of Experimental Biology* **198**, 1065–1070 (1995).

57. C. H. Greenewalt, *Hummingbirds* (Doubleday and Co., New York, 1960).

58. V. M. Ortega-Jimenez, R. Dudley, Flying in the rain: hovering performance of Anna's hummingbirds under varied precipitation. *Proc. R. Soc. B.* **279**, 3996–4002 (2012).

59. P. Chai, J. S. C. Chen, R. Dudley, Transient Hovering Performance of Hummingbirds Under Conditions of Maximal Loading. *Journal of Experimental Biology* **200**, 921–929 (1997).

60. B. Cheng, B. W. Tobalske, D. R. Powers, T. L. Hedrick, S. M. Wethington, G. T. C. Chiu, X. Deng, Flight mechanics and control of escape manoeuvres in hummingbirds I. Flight kinematics. *Journal of Experimental Biology*, jeb.137539 (2016).

61. J. Lynch, J. Gau, S. Sponberg, N. Gravish, Dimensional analysis of spring-wing systems reveals performance metrics for resonant flapping-wing flight. *J. R. Soc. Interface.* **18** (2021).

62. B. Cote, C. Casey, M. Jankauski, Wing inertia influences the phase and amplitude relationships between thorax deformation and flapping angle in bumblebees. *Bioinspir. Biomim.* **20**, 014001 (2025).

63. M. Keennon, K. Klingebiel, H. Won, "Development of the Nano Hummingbird: A Tailless Flapping Wing Micro Air Vehicle" in *50th AIAA Aerospace Sciences Meeting Including the New Horizons Forum and Aerospace Exposition* (American Institute of Aeronautics and Astronautics, Nashville, Tennessee, 2012; https://arc.aiaa.org/doi/10.2514/6.2012-588).

64. A. Roshanbin, H. Altartouri, M. Karásek, A. Preumont, COLIBRI: A hovering flapping twin-wing robot. *International Journal of Micro Air Vehicles* **9**, 270–282 (2017).

65. H. V. Phan, S. Aurecianus, T. Kang, H. C. Park, KUBeetle-S: An insect-like, tailless, hover-capable robot that can fly with a low-torque control mechanism. *International Journal of Micro Air Vehicles* **11**, 175682931986137 (2019).

66. D. Coleman, M. Benedict, V. Hrishikeshavan, I. Chopra, "Design, development and flight-testing of a robotic hummingbird" in *AHS 71st Annual Forum* (Virginia Beach, Virginia, 2015), pp. 5–7.

67. Z. Tu, F. Fei, X. Deng, Untethered Flight of an At-Scale Dual-motor Hummingbird Robot with Bio-inspired Decoupled Wings. *IEEE Robot. Autom. Lett.*, 1–1 (2020).

68. S. Xiao, Y. Shi, Z. Wang, Z. Ni, Y. Zheng, H. Deng, X. Ding, Lift system optimization for hover-capable flapping wing micro air vehicle. *Front. Mech. Eng.* **19**, 19 (2024).

69. G.-K. Lau, Y.-W. Chin, J. T.-W. Goh, R. J. Wood, Dipteran-Insect-Inspired Thoracic Mechanism With Nonlinear Stiffness to Save Inertial Power of Flapping-Wing Flight. *IEEE Trans. Robot.* **30**, 1187–1197 (2014).

70. J. M. Wakeling, C. P. Ellington, Dragonfly Flight: II. Velocities, Accelerations and Kinematics of Flapping Flight. *Journal of Experimental Biology* **200**, 557–582 (1997).

71. J. Wu, S. Chu, L. Chen, Y. Zhang, The roles of body and wing pitching angles in hovering butterflies. *Physics of Fluids* **37**, 051904 (2025).

72. J. Young, S. M. Walker, R. J. Bomphrey, G. K. Taylor, A. L. R. Thomas, Details of Insect Wing Design and Deformation Enhance Aerodynamic Function and Flight Efficiency. *Science* **325**, 1549–1552 (2009).

73. R. J. Bomphrey, T. Nakata, N. Phillips, S. M. Walker, Smart wing rotation and trailing-edge vortices enable high frequency mosquito flight. *Nature* **544**, 92–95 (2017).

74. Y. Liu, L. Liu, M. Sun, Power requirements in hovering flight of mosquitoes. *Physics of Fluids* (2024).

75. U. M. Norberg, T. H. Kunz, J. F. Steffensen, Y. Winter, O. V. Helversen, The cost of hovering and forward flight in a nectar-feeding bat, Glossophaga soricina, estimated from aerodynamic theory. *Journal of experimental biology* **182**, 207–227 (1993).

76. X. Cheng, M. Sun, Very small insects use novel wing flapping and drag principle to generate the weight-supporting vertical force.

77. G. Luo, M. Sun, The effects of corrugation and wing planform on the aerodynamic force production of sweeping model insect wings. *ACTA MECH SINICA* **21**, 531–541 (2005).

78. H. K. Kwon, J. W. Chang, Effects of shapes and kinematics of hovering flapping wings on aerodynamic forces and vortex structures. *Sci Rep* **15**, 5098 (2025).